\documentclass[aps,preprintnumbers,amsmath,amssymb,twocolumn,superscriptaddress,floatfix]{revtex4}
\usepackage{bbm}
\usepackage{color}
\usepackage{dsfont}
\usepackage{mathrsfs}
\usepackage{amsmath,amssymb,bm,comment}
\usepackage{hyperref}
\usepackage{graphicx}
\usepackage{dcolumn}
\usepackage{bm}

\begin{document}

\title{Frame Dependence in Generalized Chiral Kinetic Theory}

\author{Shu-Xiang Ma}
\affiliation{Shandong Provincial Key Laboratory of Nuclear Science, Nuclear Energy Technology and Comprehensive Utilization, Weihai Frontier Innovation Institute of Nuclear Technology, School of Nuclear Science, Energy and Power Engineering, Shandong University, Shandong 250061, China}

\author{Jian-Hua Gao}
\email{gaojh@sdu.edu.cn}
\affiliation{Shandong Provincial Key Laboratory of Nuclear Science, Nuclear Energy Technology and Comprehensive Utilization, Weihai Frontier Innovation Institute of Nuclear Technology, School of Nuclear Science, Energy and Power Engineering, Shandong University, Shandong 250061, China}

\affiliation{Weihai Research Institute of Industrial Technology of Shandong University, Weihai
264209, China}

\begin{abstract}
We investigate the frame dependence of distribution functions within the framework of generalized chiral kinetic theory. Based on the derived transformation rules governing the choice of frame, we analytically obtain the global equilibrium solution in the presence of vorticity and electromagnetic fields. Our results show that, under the assumption of a  varying electromagnetic field, these equilibrium solutions can be uniquely determined.
\end{abstract}

\maketitle

\section{Introduction}

Recently, novel chiral and spin effects have attracted significant attention in relativistic heavy-ion collisions, such as the chiral magnetic effect \cite{Vilenkin:1980fu,Kharzeev:2007jp,Fukushima:2008xe}, chiral vorticity effect \cite{Vilenkin:1978hb,Kharzeev:2007tn,Erdmenger:2008rm,Banerjee:2008th}, global polarization \cite{Liang:2004ph,Betz:2007kg,Gao:2007bc,Huang:2011ru,Becattini:2013fla,Becattini:2013vja,STAR:2017ckg}, spin alignment \cite{Liang:2004xn,Acharya:2019vpe,STAR:2022fan}, and so on.
These various chiral and spin effects have greatly stimulated theoretical research, especially in the context of quantum kinetic theory. In recent years, quantum kinetic theory has been developed in various aspects, such as extensions from massless \cite{Gao:2012ix,Stephanov:2012ki,Son:2012zy,Chen:2012ca,Manuel:2013zaa,Chen:2014cla,Chen:2015gta,Hidaka:2016yjf,Mueller:2017lzw, Huang:2018wdl,Hidaka:2018ekt,Gao:2018wmr,Gao:2018jsi,Liu:2018xip,Lin:2019fqo} to massive fermions \cite{Gao:2019znl,Weickgenannt:2019dks,Hattori:2019ahi,Wang:2019moi,Sheng:2020oqs,Guo:2020zpa,Li:2019qkf,Sheng:2022ssd,Ma:2022ins}, from Abelian to non-Abelian frameworks \cite{Mueller:2017lzw,Luo:2021uog,Yang:2021fea,Xiao-Li:2023rlp,Luo:2025cdw}, from first- to second-order formulations \cite{Gao:2018wmr,Gorbar:2017cwv,Yang:2020mtz,Hayata:2020sqz,Mameda:2023ueq,Yang:2024hmd,Yang:2024sfp}, from flat to curved spacetime \cite{Hayata:2020sqz,Liu:2020flb,Gao:2020gcf,Mameda:2022ojk}, from collisionless to collisional dynamics \cite{Chen:2015gta,Yang:2020hri,Wang:2020pej,Bhadury:2020puc,Weickgenannt:2021cuo,Fauth:2021nwe,Lin:2021mvw,Fang:2022ttm,Wang:2022yli,Weickgenannt:2024ibf,Lin:2024cxo,Wu:2024gbw}, and from fermions to gauge bosons \cite{Mameda:2022ojk,Yamamoto:2017uul,Huang:2020kik,Hattori:2020gqh,Comadran:2024ddu,Comadran:2024lqz}. Recent reviews on quantum kinetic theory can be found in Refs. \cite{Gao:2020vbh,Gao:2020pfu,Liu:2020ymh,Jiang:2021ict,Hidaka:2022dmn,Jian-Hua:2023viv}.

In quantum kinetic theory, the distribution functions can exhibit nontrivial dependence on the frame in which they are defined. For instance, in chiral kinetic theory (CKT) for massless fermions, the chiral distribution functions depend on the choice of frame \cite{Chen:2014cla,Chen:2015gta}. When transforming the distribution function from one frame to another, a nontrivial side-jump term must be included to preserve Lorentz invariance. In our previous work \cite{Ma:2022ins}, a generalized chiral kinetic theory (GCKT) for fermions of arbitrary mass was derived. This framework provides a convenient formalism for describing quantum transport of arbitrary-mass fermions and ensures a smooth transition from massive to massless cases.
However, in \cite{Ma:2022ins}, the transformation rules for distribution functions across different frames were not examined. Moreover, the global equilibrium solutions were not uniquely determined, leading to differing results in \cite{Ma:2022ins} and \cite{Gao:2019znl}. In the present work, we devote special attention to studying the frame-dependence transformation rules for distribution functions within the GCKT. Based on the derived transformation rules, we analytically obtain the global equilibrium solution in the presence of vorticity and electromagnetic fields. Our results demonstrate that, under the assumption of a  varying electromagnetic field, these equilibrium solutions can be uniquely determined.

In Section \ref{sec:Wigner}, we provide a brief review of the Wigner function formalism for massive Dirac fermions. Section \ref{sec:GCKT} presents the main results of the GCKT. In Section \ref{sec:Frame}, we derive the transformation rules governing the frame dependence of the distribution functions. Section \ref{sec:Global} is devoted to determining the Wigner functions in global equilibrium under the influence of vorticity and electromagnetic fields. In Section \ref{sec:General}, we demonstrate that these equilibrium Wigner functions can be uniquely determined under  varying electromagnetic fields. Finally, a summary of our findings is given in Section \ref{sec:summary}.

Throughout this work, we  employ the Minkowski metric convention $g^{\mu\nu}=\mathrm{diag}(1,-1,-1,-1)$ and  Levi-Civita  tensor convention $\epsilon^{0123}=1$. We adopt the natural unit with $\hbar = c =1$ unless otherwise stated..

\section{Wigner functions and Equations}
\label{sec:Wigner}
In the Wigner function framework \cite{Heinz:1983nx,Elze:1986qd,Vasak:1987um,Zhuang:1995pd}, the Wigner function $W(x,p)$ for Dirac fermions in background electromagnetic field is defined as the ensemble average of the gauge invariant Wigner operator:
\begin{eqnarray*}
\label{wigner}
 W(x,p)=\int\frac{d^4 y}{(2\pi)^4}
e^{-ip\cdot y}\left\langle \bar\psi(x_+) U(x_+,x_-)
\psi(x_-)\right\rangle,
\end{eqnarray*}
where $x_\pm\equiv x\pm y/2 $ and  $U$  denotes  the Wilson line
\begin{eqnarray*}
\label{link}
U(x_+,x_-) \equiv e^{-iy^\mu\int_0^1 ds A_\mu\left(x-\frac{1}{2}y+sy\right)}
\end{eqnarray*}
that ensures  gauge invariance. The electric charge has been absorbed into the gauge potential $A_\mu$.
The Wigner function is a matrix in spinor space and can be decomposed as
\begin{eqnarray*}
\label{decomposition}
W=\frac{1}{4}(\mathscr{F}+i\gamma^5 \mathscr{P}+\gamma^\mu \mathscr{V}_\mu +\gamma^5\gamma^\mu \mathscr{A}_\mu
+\frac{1}{2}\sigma^{\mu\nu} \mathscr{S}_{\mu\nu}).
\end{eqnarray*}
In chiral kinetic theory (CKT) or its generalized version GCKT, chiral Wigner functions are introduced as
\begin{eqnarray*}
\label{Jsmu}
\mathscr{J}^\mu_s &=& \frac{1}{2}\left(\mathscr{V}^\mu + s \mathscr{A}^\mu\right),
\end{eqnarray*}
where  $s=\pm$ denotes right-hand/left-hand chirality. With these functions,
 the Wigner equations can be organized into two groups. Group 1 reads
\begin{eqnarray*}
\label{Js-cs}
\Pi_\mu\mathscr{J}^\mu_s &=& \frac{1}{2} m  \mathscr{F},\\
\label{Js-ev1}
-\hbar \nabla_\mu\mathscr{J}^\mu_s &=&  m s\mathscr{P},\\
\label{Js-ev2}
\hbar\left(\nabla^\mu \mathscr{J}^\nu_s -\nabla^\nu \mathscr{J}^\mu_s\right)
-2s\epsilon^{\mu\nu\rho\sigma}\Pi_\rho \mathscr{J}_{s\sigma} &=& m \mathscr{S}^{\mu\nu},
\end{eqnarray*}
and Group 2 reads
\begin{eqnarray*}
\label{F-cs}
\Pi^\mu\mathscr{F}+\frac{1}{2}\hbar \nabla_\nu\mathscr{S}^{\mu\nu} &=& m \sum_{s}\mathscr{J}^\mu_{s},\\
\label{P-cs}
\Pi^\mu\mathscr{P}+\frac{1}{2}\hbar\nabla_\nu {\tilde{\mathscr{S}}}^{\mu\nu} &=&0,\\
\label{P-ev}
\hbar\nabla_\mu\mathscr{P}-2\Pi^\nu\tilde{\mathscr{S}}_{\mu\nu}&=& 2m \sum_{s} s \mathscr{J}^\mu_{s},\\
\label{F-ev}
\hbar\nabla_\mu\mathscr{F}-2\Pi^\nu\mathscr{S}_{\mu\nu}&=&0,
\end{eqnarray*}
where $\tilde{ \mathscr{S}}^{\mu\nu}=\epsilon^{\mu\nu\alpha\beta} \mathscr{S}_{\alpha\beta}/2$. In background-field approximation, the operators $\nabla^\mu$ and $\Pi^\mu$ are  defined as
\begin{eqnarray*}
\label{nabla-mu-c}
\nabla^\mu &\equiv& \partial^\mu_x- j_0\left(\frac{1}{2}\hbar\partial^p\cdot \partial_x \right)F^{\mu\nu}\partial_\nu^p,\\
\label{Pi-mu-c}
\Pi^\mu &\equiv& p^\mu -\frac{1}{2}\hbar
j_1\left(\frac{1}{2}\hbar \partial^p\cdot \partial_x\right)F^{\mu\nu}\partial_\nu^p,
\end{eqnarray*}
where $j_0(z)$ and $j_1(z)$ are  spherical Bessel functions and $\partial_x$ acts only on the field strength tensor, not on the Wigner
function. To prepare for  the semiclassical expansion in the next section, we have restored the explicit dependence on $\hbar$ in Wigner equations
or operator definitions.
In the chiral limit, Group 1 decouples from Group 2 and give rise to the chiral kinetic equation \cite{Gao:2018wmr}.

\section{Generalized chiral kinetic theory}
\label{sec:GCKT}
This section reviews the derivation of  the GCKT  to  first order in $\hbar$ and  present the derivation in  more detail and in a stepwise manner  compared with  Ref.\cite{Ma:2022ins}.
The Wigner functions can be expanded in powers of  $\hbar$ as
\begin{eqnarray*}
\label{Js-expansion}
\mathscr{J}^\mu_s &=& \mathscr{J}^{(0)\mu}_s +\hbar  \mathscr{J}^{(1)\mu}_s +\hbar^2  \mathscr{J}^{(2)\mu}_s +\cdots,
\end{eqnarray*}
with analogous expansions for the other Wigner functions $\mathscr{F}$, $\mathscr{P}$, and $\mathscr{S}_{\mu\nu}$.
At first order, the operators $\nabla^\mu $ and $\Pi^\mu$ reduce to
\begin{eqnarray*}
\nabla^\mu = \partial^\mu_x- F^{\mu\nu}\partial_\nu^p,\ \ \  \Pi^\mu = p^\mu.
\end{eqnarray*}
To disentangle the Wigner equations, we introduce  a timelike  4-vector $n^\mu$ normalized as $n^2=1$. For simplicity, $n^\mu$ is  taken to be constant and independent of $x$ or $p$. In Section \ref{sec:General}, we will show that the GCKT formulated with a general $n^\mu$ can be recovered through  the frame-dependent transformation of the distribution functions.
Using $n^\mu$, any 4-vector $X^\mu$ can be decomposed as $X^\mu=X_n n^\mu + \bar X^\mu,$
where $X_n=X\cdot n$ and $\bar X^\mu = \Delta^{\mu\nu}X_\nu$ with $\Delta^{\mu\nu}=g^{\mu\nu}-n^\mu n^\nu$.
Similarly, the antisymmetric  tensors  $F^{\mu\nu}$ and  $\mathscr{S}^{\mu\nu}$ can be expressed in terms of electric  and magnetic field,  and in terms of the electric-moment distribution function $\mathscr{K}^\mu$ and magnetic-momentum distribution function $\mathscr{M}^\mu$, respectively:
\begin{eqnarray}
F^{\mu\nu}&=&E^\mu n^\nu -E^\nu n^\mu -\bar\epsilon^{\mu\nu\sigma}B_\sigma,\nonumber\\
\label{S-KM}
\mathscr{S}^{\mu\nu}&=& \mathscr{K}^\mu n^\nu - \mathscr{K}^\nu n^\mu - \bar\epsilon^{\mu\nu\sigma}\mathscr{M}_{\sigma},
\end{eqnarray}
where we have defined the totally spacelike antisymmetric tensor $\bar\epsilon_{\mu\alpha\beta}=\epsilon_{\mu\nu\alpha\beta}n^\nu$.
The inverse relations are given by
\begin{eqnarray*}
E^\mu &=& F^{\mu\nu}n_\nu ,\ \ B^\mu=\tilde F^{\mu\nu}n_\nu,\\
\mathscr{K}^\mu &=& \mathscr{S}^{\mu\nu} n_\nu ,\ \ \mathscr{M}^\mu=\tilde{\mathscr{S}}^{\mu\nu}n_\nu,
\end{eqnarray*}
where $\tilde F^{\mu\nu}=\epsilon^{\mu\nu\alpha\beta}F_{\alpha\beta}/2$.
Substituting these expansions into the Wigner equation and extracting terms order by order, we can derive the Wigner equations at each order. It should be noted that, to obtain the GCKT to first order, we require the Wigner equations up to second order.
\subsection{ Wigner equations at zeroth order}
At zeroth order, with the decomposition via $n^\mu$, Group 1 can be rewritten as
\begin{eqnarray}
\label{Js-cs-0-nbar}
p_n\mathscr{J}^{(0)}_{sn}+\bar p_\mu\bar{\mathscr{J}}^{(0)\mu}_s &=& \frac{1}{2} m  \mathscr{F}^{(0)},\\
\label{Js-ev1-0-a}
0 &=&  m s\mathscr{P}^{(0)},\\
\label{Js-ev2-0-n}
2s\left(\bar p^\mu \mathscr{J}_{ns}^{(0)} - p_n \bar{\mathscr{J}}_{s}^{(0)\mu} \right)
&=& m{\mathscr{M}}^{(0)\mu},\\
\label{Js-ev2-0-bar}
2s\left(\bar p^\mu \bar{\mathscr{J}}_{s}^{(0)\nu} - \bar p^\nu\bar{ \mathscr{J}}_{s}^{(0)\mu} \right)
&=& m \bar\epsilon^{\mu\nu\beta} \mathscr{K}^{(0)}_\beta,
\end{eqnarray}
where the last two equations correspond to the timelike-spacelike and spacelike-spacelike components of the tensor equation
 along the direction $n^\mu$ in the last equation of Group 1, respectively.
In Group 2, all the equations are the vector equations and can be
decomposed  into timelike component and spacelike component as

\begin{eqnarray}
\label{F-cs-0-n}
p_n\mathscr{F}^{(0)} &=& m \sum_{s}\mathscr{J}^{(0)}_{sn},\\
\label{F-cs-0-bar}
\bar p^\mu\mathscr{F}^{(0)} &=& m \sum_{s}\bar{\mathscr{J}}^{(0)\mu}_{s},\\
\label{P-cs-0-n}
p_n\mathscr{P}^{(0)} &=&0,\\
\label{P-cs-0-bar}
\bar p^\mu\mathscr{P}^{(0)} &=&0,\\
\label{tS-0-n}
\bar p^\nu{\mathscr{M}}^{(0)}_{\nu}&=& m \sum_{s} s \mathscr{J}^{(0)}_{sn},\\
\label{tS-0-bar}
-p_n{\mathscr{M}}_{\mu}^{(0)} -  \bar p^\nu \bar\epsilon_{\mu\nu\beta} \mathscr{K}^{(0)\beta} &=&
 m \sum_{s} s\bar{ \mathscr{J}}^{(0)}_{s\mu},\\
\label{S-0-n}
\bar p^\nu\mathscr{K}_{\nu}^{(0)}&=&0,\\
\label{S-0-bar}
-p_n \mathscr{K}_{\mu}^{(0)} + \bar p^\nu \bar\epsilon_{\mu\nu\beta} \mathscr{M}^{(0)\beta} &=&0.
\end{eqnarray}
From Eqs. (\ref{Js-ev2-0-n}),(\ref{F-cs-0-n}),(\ref{P-cs-0-n}), (\ref{S-0-bar}), we obtain, respectively,
\begin{eqnarray}
\label{barJs-Jsn-M-0}
\bar{ \mathscr{J}}_{s}^{(0)\mu}
 &=&\frac{1}{p_n}\bar  p^\mu \mathscr{J}_{sn}^{(0)}  -\frac{s  m }{2p_n} \mathscr{M}^{(0)\mu},\\
\label{F-Jsn-0}
\mathscr{F}^{(0)} &=& \frac{m}{p_n} \sum_{s}\mathscr{J}_{sn}^{(0)},\\
\label{P-Jsn-0}
\mathscr{P}^{(0)} &=& 0,\\
\label{K-M-0}
 \mathscr{K}_{\mu}^{(0)} &=&  \frac{1}{p_n} \bar \epsilon_{\mu\alpha\beta} p^\alpha  \mathscr{M}^{(0)\beta}.
\end{eqnarray}
It is obvious, or easy to verify, that with these equations,  Eqs. (\ref{Js-ev1-0-a}),(\ref{Js-ev2-0-bar}) ,(\ref{F-cs-0-bar}), (\ref{P-cs-0-bar}),(\ref{S-0-n}),  are all satisfied automatically.
Substituting Eqs. (\ref{barJs-Jsn-M-0}) and (\ref{F-Jsn-0})  into Eq.(\ref{Js-cs-0-nbar}) along with Eq.(\ref{tS-0-n}),
we obtain the on-shell condition for $\mathscr{J}^{(0)}_{sn}$
\begin{eqnarray*}
\frac{p^2-m^2}{p_n}\mathscr{J}^{(0)}_{sn}&=&0.
\end{eqnarray*}
From this, the general solution can be written as
\begin{eqnarray}
\label{Jsn-onshell-0}
\mathscr{J}^{(0)}_{sn}&=& p_n  \mathcal{J}^{(0)}_{sn} \delta\left(p^2-m^2\right),
\end{eqnarray}
where $\mathcal{J}^{(0)}_{sn}$ is an arbritrary regular function at $p^2 = m^2$.
Substituting Eqs. (\ref{barJs-Jsn-M-0}) and (\ref{K-M-0})  into Eq.(\ref{tS-0-bar}) along with Eq.(\ref{tS-0-n}),
we obtain the on-shell condition for ${\mathscr{M}}_{\mu}^{(0)}$
\begin{eqnarray*}
\frac{p^2-m^2}{p_n}{\mathscr{M}}_{\mu}^{(0)}&=&0.
\end{eqnarray*}
From this, the general solution can be written as
\begin{eqnarray}
\label{Mmu-onshell-0}
{\mathscr{M}}_{\mu}^{(0)} &=& p_n {\mathcal{M}}_{\mu}^{(0)} \delta\left(p^2-m^2\right),
\end{eqnarray}
where  ${\mathcal{M}}_{\mu}^{(0)}$ is an arbritrary regular function at $p^2 = m^2$.
From the relation (\ref{S-KM}), the antisymmetric tensor $\mathscr{S}^{(0)\mu\nu}$ and its dual tensor $\tilde{\mathscr{S}}^{(0)\mu\nu}$ can be expressed as, respectively,
\begin{eqnarray}
\label{S-M}
\mathscr{S}^{(0)\mu\nu}&=&\frac{1}{p_n} \epsilon^{\mu\nu\alpha\beta}p_\alpha \mathscr{M}_{\beta},\nonumber\\
\label{tS-M}
\tilde{\mathscr{S}}^{(0)\mu\nu}
&=&\frac{1}{p_n}\left(\mathscr{M}^{(0)\mu} p^\nu - \mathscr{M}^{(0)\nu} p^\mu \right).
\end{eqnarray}
All these equations above indicate that we can choose  $\mathscr{J}^{(0)}_{sn}$ and ${\mathscr{M}}_{\mu}^{(0)}$ as  the basic distribution functions, since all other Wigner functions can be expressed in terms of these
functions.

The only remianing equation, Eq.(\ref{tS-0-n}), implies that the longitudinal component of the spacelike vector ${\mathscr{M}}_{\mu}^{(0)}$ along the direction $\bar p_\mu$ is not independent.  Hence, we can decompose the magnetic moments $\mathscr{M}^{\mu} $ into parts parallel  and orthogonal to the spacelike momentum $\bar p^\mu$.
\begin{eqnarray}
\label{M-onshell-0-a}
{\mathscr{M}}^{(0)\mu} &=& {\mathscr{M}}^{(0)\mu}_{\parallel} +  {\mathscr{M}}^{(0)\mu}_{\perp},\\
{\mathscr{M}}^{(0)\mu}_{\parallel}&=&  p_n{\mathcal{M}}^{(0)\mu}_{\parallel} \delta(p^2-m^2),\\
{\mathscr{M}}^{(0)\mu}_{\perp} &=&p_n {\mathcal{M}}^{(0)\mu}_{\perp} \delta(p^2-m^2),
\end{eqnarray}
with the relations
\begin{eqnarray}
\label{Mpara-0}
& &\mathcal{M}^{(0)\mu}_\parallel =\frac{m }{\bar p^2} \bar p^\mu \sum_{s} s \mathcal{J}^{(0)}_{sn},\\
& & \bar p_\mu {\mathcal{M}}^{(0)\mu}_{\perp} \delta(p^2-m^2) = 0.
\end{eqnarray}
Therefore, we can identify $\mathcal{J}^{(0)}_{sn}$ and ${\mathcal{M}}_{\perp}^{(0)\mu}$ as the final basic distribution functions.

\subsection{Wigner equations at first order}
At first order, with the decomposition via $n^\mu$, Group 1 can be rewritten as
\begin{eqnarray}
\label{Js-cs-1-nbar}
p_n\mathscr{J}^{(1)}_{sn}+\bar p_\mu\bar{\mathscr{J}}^{(1)\mu}_s &=& \frac{1}{2} m  \mathscr{F}^{(1)},\\
\label{Js-ev1-1-a}
-\nabla_\mu \mathscr{J}_s^{(0)\mu} &=&   m s\mathscr{P}^{(1)},\\
\label{Js-ev2-1-n}
2s(\bar p^\mu \mathscr{J}_{ns}^{(1)} - p_n \bar{\mathscr{J}}_{s}^{(1)\mu})
+\bar\epsilon^{\mu\rho\sigma}\nabla_\rho \mathscr{J}^{(0)}_{s\sigma}
&=& m{\mathscr{M}}^{(1)\mu},\\
\label{Js-ev2-1-bar}
2s(\bar p^\mu \bar{\mathscr{J}}_{s}^{(1)\nu} - \bar p^\nu\bar{ \mathscr{J}}_{s}^{(1)\mu})& &\nonumber\\
+\bar\epsilon^{\mu\nu\rho}n^\sigma( \nabla_\rho \mathscr{J}_{s\sigma}^{(0)}
-\nabla_\sigma \mathscr{J}_{s\rho}^{(0)})
&=& m \bar\epsilon^{\mu\nu\beta} \mathscr{K}^{(1)}_\beta,\hspace{0.7cm}
\end{eqnarray}
and Group 2  can be
decomposed  into timelike component and spacelike component as
\begin{eqnarray}
\label{F-cs-1-n}
p_n\mathscr{F}^{(1)}+\frac{1}{2}n_\mu\nabla_\nu \mathscr{S}^{(0)\mu\nu}
  &=& m \sum_{s}\mathscr{J}^{(1)}_{sn},\\
\label{F-cs-1-bar}
\bar p^\mu\mathscr{F}^{(1)} +\frac{1}{2}{\Delta^{\mu}}_\lambda\nabla_\nu \mathscr{S}^{(0)\lambda\nu}
&=& m \sum_{s}\bar{\mathscr{J}}^{(1)\mu}_{s},\\
\label{P-cs-1-n}
p_n\mathscr{P}^{(1)} +\frac{1}{2}n_\mu\nabla_\nu \tilde{\mathscr{S}}^{(0)\mu\nu} &=&0,\\
\label{P-cs-1-bar}
\bar p^\mu\mathscr{P}^{(1)}+\frac{1}{2}{\Delta^{\mu}}_\lambda\nabla_\nu\tilde{\mathscr{S}}^{(0)\lambda\nu} &=&0,\\
\label{tS-1-n}
n^\mu \nabla_\mu \mathscr{P}^{(0)}+2\bar p^\nu{\mathscr{M}}^{(1)}_{\nu}&=& 2m \sum_{s} s \mathscr{J}^{(1)}_{sn},\\
\label{tS-1-bar}
 \bar\nabla_\mu \mathscr{P}^{(0)}
-2p_n{\mathscr{M}}_{\mu}^{(1)} - 2 \bar\epsilon_{\mu\nu\beta} \bar p^\nu \mathscr{K}^{(1)\beta} &=&
 2m \sum_{s} s\bar{ \mathscr{J}}^{(1)}_{s\mu},\hspace{0.6cm}\\
\label{S-1-n}
n^\mu \nabla_\mu \mathscr{F}^{(0)}+ 2\bar p^\nu\mathscr{K}_{\nu}^{(1)}&=&0,\\
\label{S-1-bar}
\bar\nabla_\mu \mathscr{F}^{(0)}
-2p_n \mathscr{K}_{\mu}^{(1)} + 2\bar \epsilon_{\mu\nu\beta} \bar p^\nu  \mathscr{M}^{(1)\beta} &=&0.
\end{eqnarray}
From Eqs.(\ref{Js-ev2-1-n}),(\ref{F-cs-1-n}), (\ref{P-cs-1-n}),(\ref{S-1-bar}),we obtain, respectively,
\begin{eqnarray}
\label{barJs-Jsn-M-1}
\bar{ \mathscr{J}}_{s}^{(1)\mu}
 &=&\frac{\bar p^\mu}{p_n}  \mathscr{J}_{sn}^{(1)}  -\frac{s  m }{2p_n} \mathscr{M}^{(1)\mu}
 +\frac{s}{2p_n}\bar\epsilon^{\mu\rho\sigma}\nabla_\rho \mathscr{J}^{(0)}_{s\sigma},\hspace{0.6cm}\\
\label{F-Jsn-1}
\mathscr{F}^{(1)} &=& \frac{m}{p_n} \sum_{s}\mathscr{J}_{sn}^{(1)}
-\frac{1}{2p_n}n_\mu \nabla_\nu \mathscr{S}^{(0)\mu\nu},\\
\label{P-Jsn-1}
\mathscr{P}^{(1)} &=& -\frac{1}{2p_n}n_\mu \nabla_\nu \tilde{\mathscr{S}}^{(0)\mu\nu},\\
\label{K-M-1}
 \mathscr{K}_{\mu}^{(1)} &=&  \frac{1}{p_n} \bar \epsilon_{\mu\alpha\beta} p^\alpha  \mathscr{M}^{(1)\beta}
 +\frac{1}{2p_n}{\Delta_{\mu}}^\lambda \nabla_\lambda \mathscr{F}^{(0)}.
\end{eqnarray}
From Eq.(\ref{tS-1-n}), we obtain
\begin{eqnarray}
\label{M-p-1}
\bar p^\nu{\mathscr{M}}^{(1)}_{\nu}&=& m \sum_{s} s \mathscr{J}^{(1)}_{sn},
\end{eqnarray}
where we have used  Eq.(\ref{P-Jsn-0}).
Substituting Eq.(\ref{P-Jsn-1}) into Eq.(\ref{Js-ev1-1-a}) and using Eqs.(\ref{barJs-Jsn-M-0}) and (\ref{tS-M}) yields
\begin{eqnarray}
\label{Jsn-eq-0-a}
p^\mu \nabla_\mu\left( \frac{\mathscr{J}_{sn}^{(0)}}{p_n}\right)
&=&\frac{ m s}{2p_n^2}E_\mu {\mathscr{M}}^{(0)\mu},
\end{eqnarray}
which is  the generalized chiral kinetic equation (GCKE) for  $\mathscr{J}_{sn}^{(0)}$.
Similarly, substituting Eq.(\ref{P-Jsn-1}) into (\ref{P-cs-1-bar}) and using Eq.(\ref{K-M-0}) or (\ref{tS-M}) yields
the GCKE  for $\mathscr{M}^{(0)\mu}$
\begin{eqnarray}
\label{M-kin-eq-b}
p^\nu \nabla_\nu \left(\frac{\mathscr{M}^{(0)\mu}}{p_n}\right)
&=&\left(\frac{\bar p^\mu}{p_n}E^\nu - \bar\epsilon^{\mu\nu\alpha}B_\alpha\right)\frac{\mathscr{M}^{(0)}_\nu}{p_n}.
\hspace{0.2cm}
\end{eqnarray}
Inserting Eqs.(\ref{barJs-Jsn-M-1}) and (\ref{F-Jsn-1}) into Eq.(\ref{Js-cs-1-nbar}) and
Using Eqs.(\ref{M-p-1}), (\ref{barJs-Jsn-M-0}) and (\ref{S-M}), we obtain
\begin{eqnarray}
\frac{p^2-m^2}{p_n}\mathscr{J}^{(1)}_{sn}
&=&\frac{s}{2p_n}\bar\epsilon^{\mu\rho\sigma}  F_{\rho\sigma }\bar{\mathscr{J}}_{s\mu}^{(0)}.
\end{eqnarray}
From Eqs.(\ref{Jsn-onshell-0}) and (\ref{Mmu-onshell-0}), we have the general expression for $\mathscr{J}^{(1)}_{sn}$ as
\begin{eqnarray}
\label{Jsn-1-onshell}
& &\mathscr{J}^{(1)}_{sn} = p_n \mathcal{J}^{(1)}_{sn}\delta(p^2-m^2)\nonumber\\
& &-\frac{s}{2}\bar\epsilon^{\mu\rho\sigma}  F_{\rho\sigma }\left(  \bar p_\mu \mathcal{J}_{sn}^{(0)}
- \frac{s m}{2}\mathcal{M}^{(0)}_{\mu}\right)\delta'(p^2-m^2).
\end{eqnarray}
Substituting Eqs.(\ref{barJs-Jsn-M-1}) and (\ref{K-M-1}) into  Eq.(\ref{tS-1-bar}) and
using Eqs.(\ref{M-p-1}, (\ref{barJs-Jsn-M-0})and (\ref{F-Jsn-0}), we have
\begin{eqnarray}
\frac{p^2-m^2}{p_n}{\mathscr{M}}_{\mu}^{(1)}
&=& \frac{m}{2p_n}\bar\epsilon_{\mu\rho\sigma}F^{\rho\sigma} \sum_{s} \left( \frac{1}{p_n} \mathscr{J}_{sn}^{(0)} \right).
\end{eqnarray}
From Eq.(\ref{Jsn-onshell-0}), we have the general expression
\begin{eqnarray}
\label{Mmu-onshell-1}
{\mathscr{M}}_{\mu}^{(1)}
 &=&p_n {\mathcal{M}}_{\mu}^{(1)} \delta\left(p^2-m^2\right)\nonumber\\
& &-\frac{m}{2}\bar\epsilon_{\mu\rho\sigma}F^{\rho\sigma} \sum_{s}\mathcal{J}_{sn}^{(0)} \delta'(p^2-m^2).
\end{eqnarray}
From the relations (\ref{S-KM}), the antisymmetric tenor $\mathscr{S}^{\mu\nu}$ and dual tenor $\bar{\mathscr{S}}^{\mu\nu}$  to first order are given by
\begin{eqnarray}
\label{S-M-1}
\mathscr{S}^{(1)\mu\nu}
&=&\frac{1}{p_n} \epsilon^{\mu\nu\alpha\beta}p_\alpha \mathscr{M}_{\beta}^{(1)}\nonumber\\
& &+\frac{1}{2p_n}(n_\nu \bar\nabla_{\mu}- n_\mu \bar\nabla_{\nu})\mathscr{F}^{(0)},\nonumber\\
\label{tS-M-1}
\tilde{\mathscr{S}}^{(1)\mu\nu}&=&\frac{1}{p_n}(\mathscr{M}^{(1)\mu} p^\nu - \mathscr{M}^{(1)\nu} p^\mu)\nonumber\\
& &+\frac{1}{2p_n}\bar\epsilon^{\mu\nu\rho}\nabla_\rho  \mathscr{F}^{(0)}.
\end{eqnarray}
From the constraint (\ref{M-p-1}), we can decompose ${\mathscr{M}}^{(1)\mu}$ into parts parallel  and orthogonal to the  momentum $\bar p^\mu$:
\begin{eqnarray}
\label{M-onshell-1}
{\mathscr{M}}^{(1)\mu} &=& {\mathscr{M}}^{(1)\mu}_{\parallel} +  {\mathscr{M}}^{(1)\mu}_{\perp},
\end{eqnarray}
with the transverse constraint $\bar p_\mu{\mathscr{M}}^{(1)\mu}_{\perp}=0$. From the constraint (\ref{M-p-1}) and  the expression \ref{Jsn-1-onshell},
the parallel  and orthogonal parts can be given by, respectively,
\begin{eqnarray}
{\mathscr{M}}^{(1)\mu}_{\parallel}
&=&p_n\mathcal{M}^{(1)\mu}_\parallel \delta(p^2-m^2)\nonumber\\
& &-m \bar p^\mu  \frac{\bar p\cdot B}{\bar p^2}\sum_s \mathcal{J}^{(0)}_{sn} \delta'(p^2-m^2),\nonumber\\
\label{Mperp-1-onshell}
{\mathscr{M}}^{(1)\mu}_{\perp} &=&p_n {\mathcal{M}}^{(1)\mu}_{\perp} \delta(p^2-m^2)\nonumber\\
& &-m B^\mu_\perp \sum_s \mathcal{J}^{(0)}_{sn}\delta'(p^2-m^2),
\end{eqnarray}
with the relations
\begin{eqnarray}
\label{Mpara-1}
& &\mathcal{M}^{(1)\mu}_\parallel = \frac{m }{\bar p^2} \bar p^\mu \sum_{s} s \mathcal{J}^{(1)}_{sn},\\
& & \bar p_\mu {\mathcal{M}}^{(1)\mu}_{\perp} \delta(p^2-m^2)=0.
\end{eqnarray}
Under this decomposition, we have the GCKE for the transverse distribution function $\mathscr{M}^{(0)\mu}_\perp$:
\begin{eqnarray}
\label{GCKE-0-Mperp}
& &p^\nu \nabla_\nu \left(\frac{\mathscr{M}^{(0)\mu}_\perp}{p_n}\right)
=-\frac{ m p_n }{\bar p^2} E^\mu_\perp \sum_{s} s \frac{\mathscr{J}^{(0)}_{sn}}{p_n}\nonumber\\
& &-\left(\frac{p_n \bar p^\mu}{\bar p^2} E^\nu + \bar\epsilon^{\mu\nu\alpha}B_\alpha\right)\frac{\mathscr{M}^{(0)}_{\perp\nu}}{p_n}.
\end{eqnarray}
Note that we define the transverse electric and magnetic  fields  relative to the momentum $\bar p$,
\begin{eqnarray}
E^\mu_\perp \equiv E^\mu -\frac{\bar p^\mu}{\bar p^2}  E\cdot p,\ \ \ \
B^\mu_\perp \equiv B^\mu -\frac{\bar p^\mu}{\bar p^2}  B\cdot p.
\end{eqnarray}

\subsection{Second-order equations}

To obtain the GCKE for the first-order distribution functions $\mathscr{J}^{(1)}_{sn}$ and ${\mathscr{M}}_{\mu}^{(1)}$,
we require the second-order Wigner equations. Fortunately, only the following relevant equations are needed:
 \begin{eqnarray}
\label{Js-ev1-2-a}
\nabla_\mu \mathscr{J}_s^{(1)\mu} &=& -   m s\mathscr{P}^{(2)},\\
\label{P-cs-2-n}
p_n\mathscr{P}^{(2)} +\frac{1}{2}n_\mu\nabla_\nu \tilde{\mathscr{S}}^{(1)\mu\nu} &=&0,\\
\label{P-cs-2-bar}
\bar p^\mu\mathscr{P}^{(2)}+\frac{1}{2}{\Delta^{\mu}}_\lambda\nabla_\nu\tilde{\mathscr{S}}^{(1)\lambda\nu} &=&0.
\end{eqnarray}
Substituting Eq.(\ref{P-cs-2-n}) into Eq.(\ref{Js-ev1-2-a}) and using Eqs.(\ref{barJs-Jsn-M-1}), (\ref{Jsn-eq-0-a}) and (\ref{M-kin-eq-b}) yields
\begin{eqnarray}
\label{Jsn-eq-1-a}
& & p^\mu \nabla_\mu\left( \frac{\mathscr{J}_{sn}^{(1)}}{p_n}\right) =
-\frac{s}{4p_n} \bar\epsilon^{\mu\rho\sigma}(\partial^x_\lambda F_{\mu\rho}) \partial_p^\lambda
 \mathscr{J}_{s\sigma}^{(0)} \nonumber\\
& &+ \frac{ m s}{2p_n^2}E_\nu {\mathscr{M}}^{(1)\nu}
- \frac{s}{2p_n^2}\bar\epsilon^{\mu\rho\sigma} E_\mu \nabla_\rho
\mathscr{J}_{s\sigma}^{(0)},
\end{eqnarray}
which is  the GCKE for $\mathscr{J}_{sn}^{(1)}$.
Plugging Eq.(\ref{P-cs-2-n}) into (\ref{P-cs-2-bar}) and using Eq.(\ref{K-M-1}) or (\ref{tS-M-1}) yields
\begin{eqnarray}
\label{M-1-eq-b}
& &p^\nu \nabla_\nu \left(\frac{\mathscr{M}^{(1)\mu}}{p_n}\right)
=\left(\frac{\bar p^\mu}{p_n}E^\nu - \bar\epsilon^{\mu\nu\alpha}B_\alpha\right)\frac{\mathscr{M}^{(1)}_\nu}{p_n}
 \nonumber\\
& & - \frac{1}{2p_n^2} \bar\epsilon^{\mu\nu\rho}E_\nu \nabla_\rho  \mathscr{F}^{(0)}
-\frac{1}{4p_n}\bar\epsilon^{\mu\nu\rho}(\partial_\lambda^x F_{\nu\rho})\partial^\lambda_p  \mathscr{F}^{(0)}.\hspace{0.5cm}
\end{eqnarray}
This is  the GCKE for $\mathscr{M}^{(1)\mu}$, from which we can obtain the GCKE for the transverse part at first order
\begin{eqnarray}
\label{GCKE-1-Mperp}
& &p^\nu \nabla_\nu \left(\frac{\mathscr{M}^{(1)\mu}_\perp}{p_n}\right)
=-\frac{ m p_n }{\bar p^2} E^\mu_\perp \sum_{s} s \frac{\mathscr{J}^{(1)}_{sn}}{p_n}\nonumber\\
& &-\left(\frac{p_n}{\bar p^2} \bar p^\mu E^\nu +  \bar\epsilon^{\mu\nu\alpha}B_\alpha\right)
 \frac{\mathscr{M}^{(1)}_{\perp\nu}}{p_n}\nonumber\\
& & + \frac{p_\sigma}{2\bar p^2}\left(\bar p^\nu  \bar\epsilon^{\mu\rho\sigma}-\bar p^\rho \bar \epsilon^{\mu\nu\sigma}\right)
\nabla_\nu \left( \frac{1}{p_n} \nabla_\rho \mathscr{F}^{(0)}\right).
\end{eqnarray}

Let us now summarize the full set of elements in the GCKT up to first order:  The zeroth-order GCKEs are presented in Eqs. (\ref{Jsn-eq-0-a}) and (\ref{M-kin-eq-b}),
subject to constraint (\ref{tS-0-n}) and with the explicit forms given by (\ref{Jsn-onshell-0}) and (\ref{Mmu-onshell-0}).
The remaining zeroth-order Wigner functions are determined through relations (\ref{barJs-Jsn-M-0})-(\ref{K-M-0}) and
(\ref{tS-M}).
Alternatively, the zeroth-order GCKEs may be expressed equivalently as Eqs. (\ref{Jsn-eq-0-a}) and (\ref{GCKE-0-Mperp}),
using expressions (\ref{Jsn-onshell-0}) and (\ref{Mperp-1-onshell}),
while the other Wigner functions follow from relations (\ref{barJs-Jsn-M-0})-(\ref{K-M-0}) and
(\ref{tS-M}) together with (\ref{Mpara-0}).
At first order, the GCKEs are given by Eqs. (\ref{Jsn-eq-1-a}) and (\ref{M-1-eq-b}), constrained by (\ref{M-p-1})
and with the functional forms specified in (\ref{Jsn-1-onshell}) and (\ref{Mmu-onshell-1}).
The corresponding first-order Wigner functions are obtained from relations (\ref{barJs-Jsn-M-1})-(\ref{K-M-1}) and (\ref{tS-M-1}).
An equivalent formulation of the first-order GCKEs is provided by Eqs. (\ref{Jsn-eq-1-a}) and (\ref{GCKE-1-Mperp}),
along with expressions (\ref{Jsn-1-onshell}) and (\ref{Mperp-1-onshell}).
In this case, the remaining first-order Wigner functions are derived from relations (\ref{barJs-Jsn-M-1})-(\ref{K-M-1})
 and (\ref{tS-M-1}) in combination with (\ref{Mpara-1}).

\section{Frame dependence in the GCKT}
\label{sec:Frame}

In the GCKT, the basic distribution functions $ \mathscr{J}_{sn}$ and $ \mathscr{M}^{\mu}$ are defined with respect to  the auxiliary timelike vector $n^\mu$ and consequently they  depend on this time-like vector $n^\mu$. Since we can identify the normalized timelike vector $n^\mu$ as the velocity of a frame, it is important to discuss how these distribution functions transform between
 different frames.   We assume that the distribution functions
are defined in two different frames, $n_\mu$ and $n'_\mu$. Then the Wigner functions $\mathscr{J}_{s}^{\mu}$  and
${\mathscr{S}}^{\mu\nu}$ (or $\tilde{\mathscr{S}}^{\mu\nu}$) should not depend on the choice of $n_\mu$ or $n'_\mu$.
This condition  determines the frame dependence of distribution functions.

At the zeroth order, the independence of  Wigner functions $\mathscr{J}_{s}^{(0)\mu}$  on the choice of $n_\mu$ or $n'_\mu$,  together with the expression (\ref{barJs-Jsn-M-0}),  yields
\begin{eqnarray}
\label{Jmu-0-n-np}
\frac{p^\mu \mathscr{J}_{sn}^{(0)}}{p_n}  -\frac{s  m  \mathscr{M}_n^{(0)\mu} }{2p_n}
 =\frac{p^\mu \mathscr{J}_{sn'}^{(0)} }{p_{n'}} -\frac{s  m\mathscr{M}_{n'}^{(0)\mu} }{2p_{n'}} ,
\end{eqnarray}
while  the independence of the  Wigner functions $\tilde{\mathscr{S}}^{(0)\mu\nu}$ on the choice of $n_\mu$ or $n'_\mu$, together with the expression (\ref{tS-M}), yields
\begin{eqnarray}
\label{tS-0-n-np}
 \frac{\mathscr{M}_n^{(0)\mu} p^\nu}{p_n} -  \frac{\mathscr{M}_n^{(0)\nu} p^\mu}{p_n}
= \frac{\mathscr{M}_{n'}^{(0)\mu} p^\nu}{p_{n'}} -  \frac{\mathscr{M}_{n'}^{(0)\nu} p^\mu}{p_{n'}}.
\end{eqnarray}
Contracting both sides of Eqs. (\ref{Jmu-0-n-np}) and (\ref{tS-0-n-np}) with a four-vector $\eta^\mu$. If  $\eta\cdot p\neq 0$, we can divide both sides by $\eta\cdot p$ and obtain, respectively,
\begin{eqnarray*}
\label{Jmu-0-n-np-a}
\frac{\mathscr{J}_{sn}^{(0)} }{p_n}  -\frac{s  m (\eta\cdot  \mathscr{M}_n^{(0)})}{2p_n(\eta\cdot p)}
 &=&\frac{\mathscr{J}_{sn'}^{(0)}}{p_{n'}}   - \frac{sm(\eta\cdot \mathscr{M}_{n'}^{(0)})}
 {2p_{n'}(\eta\cdot p) },\\
\label{tS-0-n-np-a}
\frac{\mathscr{M}_n^{(0)\mu}}{p_n}- \frac{p^\mu (\eta\cdot \mathscr{M}_n^{(0)})}{p_n(\eta\cdot p)}
&=&\frac{\mathscr{M}_{n'}^{(0)\mu} }{p_{n'}} - \frac{p^\mu(\eta\cdot \mathscr{M}_{n'}^{(0)})}{p_{n'}(\eta\cdot p)}.\hspace{0.5cm}
\end{eqnarray*}
These relations express the frame dependence of the zeroth-order distribution functions $\mathscr{J}_{sn}^{(0)}$ and
$\mathscr{M}_n^{(0)\mu}$ in the GCKT.

At first-order, the independence of the first-order Wigner functions $\mathscr{J}_{s}^{(1)\mu}$  and
$\tilde{\mathscr{S}}^{(1)\mu\nu}$ on the choice of $n_\mu$ or $n'_\mu$,  together with the expressions (\ref{barJs-Jsn-M-1}) and  (\ref{tS-M-1}),  leads to, respectively,
\begin{eqnarray*}
& &\frac{p^\mu \mathscr{J}_{sn}^{(1)}}{p_n}   -\frac{s  m }{2p_n} \mathscr{M}_n^{(1)\mu}
 +\frac{s\epsilon^{\mu\nu\rho\sigma}n_\nu}{2p_n}\nabla_\rho \mathscr{J}^{(0)}_{s\sigma}\nonumber\\
 &=&\frac{p^\mu\mathscr{J}_{sn'}^{(1)} }{p_{n'}}   -\frac{s  m }{2p_{n'}} \mathscr{M}_{n'}^{(1)\mu}
 +\frac{s\epsilon^{\mu\nu\rho\sigma}n'_\nu}{2p_{n'}}\nabla_\rho \mathscr{J}^{(0)}_{s\sigma},\\
& &\frac{\mathscr{M}_n^{(1)\mu} p^\nu}{p_n} - \frac{\mathscr{M}_n^{(1)\nu} p^\mu}{p_n}
+\frac{\epsilon^{\mu\sigma\nu\rho}n_\sigma}{2p_n}\nabla_\rho  \mathscr{F}^{(0)}\nonumber\\
&=&\frac{\mathscr{M}_{n'}^{(1)\mu} p^\nu}{p_{n'}} -\frac{\mathscr{M}_{n'}^{(1)\nu} p^\mu}{p_{n'}}
+\frac{\epsilon^{\mu\sigma\nu\rho}n'_\sigma}{2p_{n'}}\nabla_\rho  \mathscr{F}^{(0)}.
\end{eqnarray*}
Similar to the zeroth order, contracting both sides with $\eta^\mu$ and dividing by $\eta\cdot p$ leads to
 the frame dependence of the first-order distribution functions $\mathscr{J}_{sn}^{(1)}$ and
$\mathscr{M}_n^{(1)\mu}$ in the GCKT:
\begin{eqnarray}
\label{Jsn-Frame-1}
& &\frac{\mathscr{J}_{sn}^{(1)}}{p_n}    - \frac{s  m(\eta\cdot  \mathscr{M}_n^{(1)})}{2p_n(\eta\cdot p)}
 + \frac{s\epsilon^{\mu\nu\rho\sigma}\eta_\mu n_\nu}{2p_n(\eta\cdot p)}\nabla_\rho \mathscr{J}^{(0)}_{s\sigma}\nonumber\\
 &=&\frac{\mathscr{J}_{sn'}^{(1)}}{p_{n'}}  - \frac{s  m (\eta\cdot \mathscr{M}_{n'}^{(1)})}{2p_{n'}(\eta\cdot p)}
 +\frac{s\epsilon^{\mu\nu\rho\sigma}\eta_\mu n'_\nu}{2p_{n'}(\eta\cdot p)}\nabla_\rho \mathscr{J}^{(0)}_{s\sigma},\ \ \ \\
\label{Mmu-Frame-1}
& &\frac{\mathscr{M}_n^{(1)\mu}}{p_n}  - \frac{p^\mu(\eta\cdot  \mathscr{M}_n^{(1)})}{p_n(\eta\cdot p)}
+ \frac{\epsilon^{\mu\sigma\nu\rho} \eta_\nu n_\sigma}{2p_n(\eta\cdot p)}\nabla_\rho  \mathscr{F}^{(0)}\nonumber\\
&=&\frac{\mathscr{M}_{n'}^{(1)\mu}}{p_{n'}} - \frac{p^\mu(\eta\cdot \mathscr{M}_{n'}^{(1)})}{p_{n'}(\eta\cdot p)}
+ \frac{\epsilon^{\mu\sigma\nu\rho}\eta_\nu n'_\sigma}{2p_{n'}(\eta\cdot p)}\nabla_\rho  \mathscr{F}^{(0)}.
\end{eqnarray}

It might seem  confusing that to discuss the frame dependence from the auxiliary vector $n^\mu$, we have introduced another
auxiliary vector $\eta^\mu$. In fact, we can avoid this complexity by choosing $\eta^\mu=p^\mu$ or $\eta^\mu=n^\mu$ or
$\eta^\mu=n^{\prime\mu}$. When we choose $\eta^\mu=p^\mu$, we find that the $p^2$ term will contribute  $m^2$ in the denominator
and lead to non-trivial chiral limit as $m=0$, which would undermine the advantage of the GCKT. When we choose $\eta^\mu=n^\mu$,
the zeroth-order transformation rules for the frame dependence are given by
\begin{eqnarray}
\label{Jmu-0-n-np-b}
\frac{\mathscr{J}_{sn'}^{(0)}}{p_{n'}}  - \frac{\mathscr{J}_{sn}^{(0)} }{p_n}
 &=&\frac{s  m} {2p_{n'} p_n } n\cdot \mathscr{M}_{n'}^{(0)},\\
\label{tS-0-n-np-b}
\frac{\mathscr{M}_{n'}^{(0)\mu} }{p_{n'}} - \frac{\mathscr{M}_n^{(0)\mu}}{p_n}
&=& \frac{ p^\mu }{p_{n'} p_n}n\cdot \mathscr{M}_{n'}^{(0)},
\end{eqnarray}
 and the first-order transformation rules are given by
\begin{eqnarray}
\label{Jmu-1-n-np-b}
\frac{\mathscr{J}_{sn'}^{(1)}}{p_{n'}} - \frac{\mathscr{J}_{sn}^{(1)}}{p_n}
 &=&\frac{s  m}{2p_{n'}  p_n}n\cdot \mathscr{M}_{n'}^{(1)}\nonumber\\
& &- \frac{s \epsilon^{\mu\nu\rho\sigma}n_\mu n'_\nu}{2p_{n'} p_n}\nabla_\rho \mathscr{J}^{(0)}_{s\sigma},\\
\label{tS-1-n-np-b}
\frac{\mathscr{M}_{n'}^{(1)\mu}}{p_{n'}} - \frac{\mathscr{M}_n^{(1)\mu}}{p_n}
&=& \frac{p^\mu }{p_{n'} p_n}n\cdot \mathscr{M}_{n'}^{(1)}\nonumber\\
& &-\frac{\epsilon^{\mu\sigma\nu\rho}n_\nu n'_\sigma}{2p_{n'} p_n}\nabla_\rho  \mathscr{F}^{(0)}.
\end{eqnarray}
The last term on the right-hand of Eq.(\ref{Jmu-1-n-np-b}) is just the side-jump term in the CKT. In the GCKT,
the first term on the right-hand side of Eq.(\ref{Jmu-1-n-np-b}) is additional term due to finite mass. Such additional terms
even exist in Eq.(\ref{Jmu-0-n-np-b}) at zeroth order.  We can regard the last term on the right-hand side of Eq.(\ref{tS-1-n-np-b}) as  the side-jump term for the distribution function $\mathscr{M}^{(1)\mu}$.  In the following sections, we will find that
the arbitrariness of auxiliary vector $\eta^\mu$ make it possible to formulate the GCKT with an arbitrary $n^\mu$ depending on spacetime, which will be used to determine the Wigner functions in global equilibrium.

\section{Wigner functions in global equilibrium}
\label{sec:Global}
In this section, we will apply  the results given in Section \ref{sec:GCKT} to find the solution for the Wigner functions in global equilibrium
under vorticity and electromagnetic fields. The zeroth-order Wigner function is obtained from free quantum field theory and provided as a given input. When the system is not polarized, the fundamental distribution functions in GCKT read
\begin{eqnarray}
\label{Jsn-0-eq}
 {\mathcal{J}}_{sn}^{(0)} &=& \frac{1}{4\pi^3}
\left[\frac{\theta (p_0-m)  }{1+ e^{\beta\cdot p-\bar\mu }}
+ \frac{\theta (-p_0-m) }{1+ e^{ -\beta\cdot p + \bar\mu } }\right],\\
\label{Mmu-0-eq}
{\mathcal{M}}^{(0)\mu}_\perp &=& 0,
\end{eqnarray}
where $\beta^\mu =  u^\mu / T  $ and $\bar \mu = \mu/T$ with temperature $T$, fluid velocity $ u^\mu $, and chemical potential $\mu$.
These expressions are the specific solutions of the GCKT at zeroth order if the following constraints in global equilibrium are satisfied  under varying  $F^{\mu\nu}$
\begin{eqnarray*}
\partial_\mu \beta_\nu + \partial_\nu \beta_\mu =0, \ \ \ \ \partial_\mu \bar\mu + F_{\mu\nu} \beta^\nu = 0.
\end{eqnarray*}
The first constraint implies that the  thermal vorticity
\begin{eqnarray*}
\Omega_{\mu\nu} & = & \frac{1}{2}(\partial_{\mu}\beta_{\nu}-\partial_{\nu}\beta_{\mu}).
\end{eqnarray*}
is a constant tensor, i.e., $\partial_{\rho}\Omega_{\mu\nu}=0$, in global equilibrium \cite{Yang:2020mtz}. For
the second constraint, applying the partial derivative $\partial_{\nu}$
on both sides and using the commutativity of partial derivatives leads to the integrability condition:
\begin{eqnarray*}
F_{\mu\lambda}\Omega_{\nu}^{\;\lambda}-F_{\nu\lambda}\Omega_{\mu}^{\;\lambda}  =  -\beta^{\lambda}(\partial_{\lambda}F_{\mu\nu}),
\end{eqnarray*}
Contracting both sides with $\epsilon^{\alpha\beta\mu\nu}/2$ gives rise to
\begin{eqnarray}
\label{con-tF-a}
{F^\alpha}_{\lambda} \tilde\Omega^{\beta\lambda}-{F^\beta}_{\lambda} \tilde\Omega^{\alpha\lambda}
&=&{\tilde{F}^{\alpha}}_{\ \ \lambda} {\Omega^{\beta\lambda}}  - {\tilde{F}^{\beta}}_{\ \ \lambda} \Omega^{\alpha\lambda}\nonumber\\
&=&  -\beta^{\lambda}(\partial_{\lambda}\tilde{F}^{\alpha\beta}).
\end{eqnarray}
The term in the last line survives only for varying electromagnetic fields and vanishes for constant ones.
As will be demonstrated, the unique determination of a global equilibrium solution is only possible when varying electromagnetic fields are present --- a condition unmet in the constant-field case.

Now, we proceed to determine the first-order solution based on the zeroth-order solutions provided in Eqs. (\ref{Jsn-0-eq}) and (\ref{Mmu-0-eq}).
As expressed in Eqs. (\ref{Jsn-1-onshell}) and (\ref{Mperp-1-onshell}), the first-order GCKEs take the form:
\begin{eqnarray}
\label{Jsn-kinetic-eq-1-cb}
& &p^\mu \nabla_\mu\left[  \mathcal{J}^{(1)}_{sn}\delta(p^2-m^2)
- \frac{s(B\cdot  \bar p)}{p_n} \mathcal{J}_{sn}^{(0)}\delta'(p^2-m^2)\right]\nonumber\\
&=&\frac{ m s}{2p_n}E_\mu
\left\{ \frac{m \bar p^\mu}{\bar p^2}\sum_{s'}  \left[ s'\mathcal{J}^{(1)}_{s'n}\delta(p^2-m^2)\right.\right.\nonumber\\
& &\left.\left.- \frac{(B\cdot  \bar p)}{p_n} \mathcal{J}_{s'n}^{(0)}\delta'(p^2-m^2)\right]
+ {\mathcal{M}}^{(1)\mu}_{\perp} \delta(p^2-m^2)\right.\nonumber\\
& &\left. - \frac{m}{p_n} B^\mu_\perp \sum_{s'} \mathcal{J}^{(0)}_{s'n}\delta'(p^2-m^2)\right\}\nonumber\\
& &-\frac{s}{2}\bar\epsilon^{\mu\rho\sigma}\nabla_\mu \left\{\frac{1}{p_n}\nabla_\rho
\left[ p_\sigma\mathcal{J}_{sn}^{(0)}\delta(p^2-m^2) \right]\right\},
\end{eqnarray}
\begin{eqnarray}
\label{M-kinetic-eq-1-cb}
& &p^\nu \nabla_\nu \left[{\mathcal{M}}^{(1)\mu}_{\perp} \delta(p^2-m^2)
- \frac{m B^\mu_\perp}{p_n}  \sum_{s'} \mathcal{J}^{(0)}_{s'n}\delta'(p^2-m^2)\right]\nonumber\\
&=&-\left(\frac{p_n}{\bar p^2} \bar p^\mu E^\nu +  \bar\epsilon^{\mu\nu\alpha}B_\alpha\right)
\left[{\mathcal{M}}^{(1)}_{\perp\nu} \delta(p^2-m^2)\right.\nonumber\\
& &\left. - \frac{m}{p_n} B_{\perp\nu} \sum_{s'} \mathcal{J}^{(0)}_{s'n}\delta'(p^2-m^2)\right]
-\frac{ m p_n }{\bar p^2}E^\mu_\perp \sum_{s'}  \nonumber\\
& &\left[ s'\mathcal{J}^{(1)}_{s'n}\delta(p^2-m^2)
- \frac{(B\cdot  \bar p)}{p_n} \mathcal{J}_{s'n}^{(0)}\delta'(p^2-m^2)\right]\nonumber\\
& & + \frac{m}{2\bar p^2}\left(\bar p^\nu  \bar\epsilon^{\mu\rho\sigma}-\bar p^\rho \bar \epsilon^{\mu\nu\sigma}\right)p_\sigma
\nabla_\nu \left\{\frac{1}{p_n} \nabla_\rho\sum_{s}\right.\nonumber\\
& &\left. \left[\mathcal{J}_{sn}^{(0)}\delta(p^2-m^2)  \right]  \right\}.
\end{eqnarray}

Let us first deal with the GCKE for $\mathcal{J}^{(1)}_{sn}$ first.
Moving the second term in the square bracket in the first line
of Eq.(\ref{Jsn-kinetic-eq-1-cb}) to the right-hand side of the equal sign yields

\begin{eqnarray*}
\label{Jsn-kinetic-eq-1-cd}
& &p^\mu \nabla_\mu\left[ \mathcal{J}^{(1)}_{sn}\delta(p^2-m^2)\right]\nonumber\\
&=&\frac{ m s}{2p_n}E_\mu
\left[ \frac{m \bar p^\mu}{\bar p^2}\sum_{s'}  s'\mathcal{J}^{(1)}_{s'n}
+ {\mathcal{M}}^{(1)\mu}_{\perp}\right] \delta(p^2-m^2)\nonumber\\
& &-\frac{ m^2 s}{2p_n^2}(E\cdot B) \sum_{s'} \mathcal{J}^{(0)}_{s'n}\delta'(p^2-m^2)\nonumber\\
& &-\frac{s}{2p_n^2}\bar\epsilon^{\mu\rho\sigma}E_\mu \nabla_\rho
\left[ p_\sigma\mathcal{J}_{sn}^{(0)}\delta(p^2-m^2) \right]\nonumber\\
& & + p^\mu \nabla_\mu\left[  \frac{s(B\cdot  \bar p)}{p_n} \mathcal{J}_{sn}^{(0)}\delta'(p^2-m^2)\right]\nonumber\\
& &{-\frac{s}{2p_n}(\partial^x_\lambda  B^\sigma)\partial_p^\lambda
\left[ p_\sigma\mathcal{J}_{sn}^{(0)}\delta(p^2-m^2) \right]}.
\end{eqnarray*}
Using the identities
\begin{eqnarray*}
 \nabla_\rho\mathcal{J}_{sn}^{(0)}&=&\mathcal{J}_{sn}^{(0)\prime} \Omega_{\rho\sigma}p^\sigma
=-\frac{1}{2} \mathcal{J}_{sn}^{(0)\prime} \epsilon_{\rho\lambda\tau\kappa}p^\lambda \tilde\Omega^{\tau\kappa},\nonumber\\
\mathcal{J}_{sn}^{(0)\prime}& \equiv &\frac{\partial \mathcal{J}_{sn}^{(0)}}{\partial (\beta\cdot p)},
\end{eqnarray*}
and the Maxwell's equation $\partial_\mu\tilde{F}^{\mu\nu}=0$, we  obtain
\begin{eqnarray}
\label{Jsn-kinetic-eq-1-ce}
& &p^\mu \nabla_\mu\left[ \mathcal{J}^{(1)}_{sn}\delta(p^2-m^2)\right]\nonumber\\
&=&\frac{ m s}{2p_n}E_\mu
\left( \frac{m \bar p^\mu}{\bar p^2}\sum_{s'} s'\mathcal{J}^{(1)}_{s'n}
+ {\mathcal{M}}^{(1)\mu}_{\perp}\right)\delta(p^2-m^2)\nonumber\\
& &+\frac{s}{2p_n^2}E_\mu \left(\bar p^\mu \tilde\Omega^{\nu\lambda}\bar p_\nu n_\lambda
-\bar p^2 \tilde\Omega^{\mu\lambda} n_\lambda\right.\nonumber\\
& &\left.\hspace{1.5cm}+ p_n \tilde\Omega^{\mu\nu}\bar p_\nu  \right) \mathcal{J}_{sn}^{(0)\prime} \delta(p^2-m^2)\nonumber\\
& &{ -\frac{s}{2p_n}\left[\beta^\lambda \partial^x_\lambda ( B\cdot p)\right]\mathcal{J}_{sn}^{(0)\prime}
\delta(p^2-m^2) }.
\end{eqnarray}
Using the constraint equation (\ref{con-tF-a}), we have
\begin{eqnarray}
\label{con-s}
\beta^{\lambda}\partial_{\lambda} B_{\mu}
&=&-\bar\epsilon_{\mu\nu\alpha} B^\alpha \tilde\Omega^{\nu\rho}n_\rho
- E^\nu \Delta_\mu^\rho\tilde\Omega_{\rho\nu}.
\end{eqnarray}
Together with  the decomposition
\begin{eqnarray}
\nabla_\mu &=& \partial_\mu^x - E_\mu \partial_{p_n} + n_\mu E^\nu \bar\partial_\nu^p -\bar \epsilon_{\mu\nu\rho} B^\nu \bar \partial^\rho_{p},
\end{eqnarray}
we can write the Eq.(\ref{Jsn-kinetic-eq-1-ce}) as
\begin{eqnarray}
& &p^\mu \nabla_\mu\left[ \mathcal{J}^{(1)}_{sn}\delta(p^2-m^2)\right]\nonumber\\
&=&\frac{ m s}{2p_n}E_\mu
\left[ \frac{m \bar p^\mu}{\bar p^2}\sum_{s'} s'\mathcal{J}^{(1)}_{s'n}
+ {\mathcal{M}}^{(1)\mu}_{\perp}\right]\delta(p^2-m^2)\\
& &+\frac{s}{2p_n^2}E_\mu \left(\bar p^\mu \tilde\Omega^{\nu\lambda}\bar p_\nu
-\bar p^2 \tilde\Omega^{\mu\lambda}  \right) n_\lambda \mathcal{J}_{sn}^{(0)\prime} \delta(p^2-m^2)\nonumber\\
& &{ +\frac{s}{2p_n}\bar\epsilon_{\mu\nu\alpha}\bar p^\mu B^\alpha \tilde\Omega^{\nu\lambda}n_\lambda\mathcal{J}_{sn}^{(0)\prime}
\delta(p^2-m^2) }.
\end{eqnarray}
Using the identity
\begin{eqnarray}
& & p^\mu \nabla_\mu
\left[ \frac{s}{2p_n} \tilde\Omega^{\nu\lambda}p_\nu n_\lambda\mathcal{J}_{sn}^{(0)\prime} \delta(p^2-m^2)\right]\nonumber\\
&=&\frac{sE_\mu}{2p_n^2} \left(\bar p^\mu \tilde\Omega^{\nu\lambda}\bar p_\nu
+ p_n^2 \tilde\Omega^{\mu\lambda}  \right) n_\lambda \mathcal{J}_{sn}^{(0)\prime} \delta(p^2-m^2)\nonumber\\
& &+\frac{s}{2p_n}\bar\epsilon_{\mu\nu\alpha}\bar p^\mu B^\alpha \tilde\Omega^{\nu\lambda}n_\lambda\mathcal{J}_{sn}^{(0)\prime}
\delta(p^2-m^2),
\end{eqnarray}
along with  a little reorganization, we obtain
\begin{eqnarray}
& &p^\mu \nabla_\mu\left[ \left(\mathcal{J}^{(1)}_{sn}-\delta\mathcal{J}^{(1)}_{sn} \right)\delta(p^2-m^2)\right]\nonumber\\
&=&\frac{ m s}{2p_n}E_\mu
\left[ \frac{m \bar p^\mu}{\bar p^2}\sum_{s'} s'\left(\mathcal{J}^{(1)}_{s'n}-\delta\mathcal{J}^{(1)}_{sn} \right) \right]\delta(p^2-m^2)\nonumber\\
& &+ \frac{ m s}{2p_n}E_\mu
\left[{\mathcal{M}}^{(1)\mu}_{\perp}-\delta{\mathcal{M}}^{(1)\mu}_{\perp} \right]\delta(p^2-m^2),
\end{eqnarray}
where we have defined
\begin{eqnarray}
\delta \mathcal{J}^{(1)}_{sn} &\equiv & \frac{s}{2p_n} \tilde\Omega^{\nu\lambda}p_\nu n_\lambda\mathcal{J}_{sn}^{(0)\prime},\\
\delta{\mathcal{M}}^{(1)\mu}_{\perp}&\equiv& \frac{m}{2p_n}(\tilde\Omega^{\mu\lambda} n_\lambda
-\frac{\bar p^\mu}{\bar p^2} \tilde\Omega^{\nu\lambda}\bar p_\nu n_\lambda)\sum_s \mathcal{J}_{sn}^{(0)\prime}.
\end{eqnarray}
Similarly, let us consider the GCKE for ${\mathcal{M}}^{(1)\mu}_{\perp}$.   Moving the second term in square bracket in the first line of Eq.(\ref{M-kinetic-eq-1-cb}) to the right-hand side of the equal sign yields

\begin{eqnarray*}
\label{M-t-EQ}
& &p^\nu \nabla_\nu \left[ {\mathcal{M}}^{(1)\mu}_{\perp} \delta(p^2-m^2) \right]\nonumber\\
&=&-\left(\frac{p_n}{\bar p^2} \bar p^\mu E^\nu +  \bar\epsilon^{\mu\nu\alpha}B_\alpha\right)
{\mathcal{M}}^{(1)}_{\perp\nu} \delta(p^2-m^2)\nonumber\\
& &-\frac{ m p_n }{\bar p^2}E^\mu_\perp \sum_{s'} s'\mathcal{J}^{(1)}_{s'n}\delta(p^2-m^2)\nonumber\\
& & - \frac{m}{2 p_n}E^\nu( \frac{\bar p^\mu \bar p^\rho}{\bar p^2}\tilde\Omega_{\nu\rho}
+  \Delta^{\mu\lambda} \tilde \Omega_{\lambda\nu}) \sum_{s'} \mathcal{J}_{sn}^{(0)\prime}  \delta(p^2-m^2)\nonumber\\
& & - \frac{m E\cdot \bar p}{2 p_n^2}(\frac{\bar p^\mu\bar p_\nu}{\bar p^2} \tilde\Omega^{\nu\rho} n_\rho
-\tilde \Omega^{\mu\nu}n_\nu)
  \sum_{s'} \mathcal{J}_{sn}^{(0)\prime}  \delta(p^2-m^2)\nonumber\\
& &{ + \frac{m p_\sigma \beta^\lambda}{2\bar p^2p_n} \left(\bar p^\mu  \partial_\lambda B^\sigma - \bar p^\sigma \partial_\lambda B^\mu \right)
 \sum_{s} \mathcal{J}_{sn}^{(0)\prime}\delta(p^2-m^2)  }.\hspace{0.7cm}
\end{eqnarray*}
Using the constraint equation (\ref{con-s}), we have
\begin{eqnarray*}
& &p^\nu \nabla_\nu \left[ {\mathcal{M}}^{(1)\mu}_{\perp} \delta(p^2-m^2) \right]\nonumber\\
&=&-(\frac{p_n}{\bar p^2} \bar p^\mu E^\nu +  \bar\epsilon^{\mu\nu\alpha}B_\alpha)
{\mathcal{M}}^{(1)}_{\perp\nu} \delta(p^2-m^2)\nonumber\\
& &-\frac{ m p_n }{\bar p^2}E^\mu_\perp \sum_{s'} s'\mathcal{J}^{(1)}_{s'n}\delta(p^2-m^2)\nonumber\\
& & - \frac{m  E\cdot \bar p }{2 p_n^2}
(\frac{\bar p^\mu \bar p_\rho }{\bar p^2}\tilde\Omega^{\rho\nu} -\tilde \Omega^{\mu\nu})n_\nu
  \sum_{s'} \mathcal{J}_{sn}^{(0)\prime}  \delta(p^2-m^2)\nonumber\\
& &{ + \frac{m B_\alpha}{2p_n}
(\bar\epsilon^{\mu\nu\alpha} - \frac{\bar p^\mu \bar p_\sigma}{\bar p^2} \bar\epsilon^{\sigma\nu\alpha})
 \tilde\Omega_{\nu\rho}n^\rho
 \sum_{s} \mathcal{J}_{sn}^{(0)\prime}\delta(p^2-m^2)  }.\hspace{0.7cm}
\end{eqnarray*}
Using the identity
\begin{eqnarray}
& &p^\nu \nabla_\nu \left[ \frac{m n_\lambda}{2p_n}( \tilde\Omega^{\mu\lambda}
-\frac{\bar p^\mu}{\bar p^2} \tilde\Omega^{\rho\lambda}\bar p_\rho )
 \sum_{s}\mathcal{J}_{sn}^{(0)\prime} \delta(p^2-m^2)\right]\nonumber\\
&=&  \frac{m}{2p_n^2} (E\cdot p) \tilde\Omega^{\mu\lambda} n_\lambda
 \sum_{s} \mathcal{J}_{sn}^{(0)\prime}\delta(p^2-m^2)
\nonumber\\
& & -\frac{m}{2\bar p^2}\left( E^\mu \bar p_\rho + \bar p^\mu E_\rho \right) \tilde\Omega^{\rho\lambda} n_\lambda
 \sum_{s}\mathcal{J}_{sn}^{(0)\prime}\delta(p^2-m^2)\nonumber\\
& & +\frac{m}{2p_n\bar p^2}\bar\epsilon^{\nu\alpha\mu}\bar p_\nu B_\alpha
\tilde\Omega^{\rho\lambda}\bar p_\rho n_\lambda \sum_{s} \mathcal{J}_{sn}^{(0)\prime}\delta(p^2-m^2)\nonumber\\
& &+\frac{m}{2p_n\bar p^2} \bar p^\mu \bar\epsilon^{\nu\alpha\rho}\bar p_\nu B_\alpha  \tilde\Omega_{\rho\lambda} n^\lambda \sum_{s}\mathcal{J}_{sn}^{(0)\prime}\delta(p^2-m^2),
\end{eqnarray}
we can write the equation as
\begin{eqnarray*}
& &p^\nu \nabla_\nu \left[\left( {\mathcal{M}}^{(1)\mu}_{\perp} - \delta {\mathcal{M}}^{(1)\mu}_{\perp} \right) \delta(p^2-m^2) \right]\nonumber\\
&=&-\left(\frac{p_n}{\bar p^2} \bar p^\mu E^\nu_\perp +  \bar\epsilon^{\mu\nu\alpha}B_\alpha\right)
\left[{\mathcal{M}}^{(1)}_{\perp\nu} -\delta{\mathcal{M}}^{(1)}_{\perp\nu} \right] \delta(p^2-m^2)\nonumber\\
& &-\frac{ m p_n }{\bar p^2}E^\mu_\perp \sum_{s} s\left(\mathcal{J}^{(1)}_{sn}
-\delta\mathcal{J}^{(1)}_{sn} \right)\delta(p^2-m^2).
\end{eqnarray*}
If we decompose the distribution functions as follows:
\begin{eqnarray}
\mathcal{J}^{(1)}_{sn}&=& J^{(1)}_{sn} + \delta\mathcal{J}^{(1)}_{sn},\\
 {\mathcal{M}}^{(1)\mu}_{\perp}
&=&{M}^{(1)\mu}_{\perp}  + \delta {\mathcal{M}}^{(1)\mu}_{\perp},
\end{eqnarray}
we can obtain the kinetic equations for ${J}^{(1)}_{sn}$ and ${{M}}^{(1)\mu}_{\perp} $
\begin{eqnarray}
\label{Jsn-1-reduce}
& &p^\mu \nabla_\mu\left[{J}^{(1)}_{sn}\delta(p^2-m^2)\right]\nonumber\\
&=&\frac{ m s E_\mu}{2p_n}
(\frac{m \bar p^\mu}{\bar p^2}\sum_{s'} s'{J}^{(1)}_{s'n} + {M}^{(1)\mu}_{\perp} )\delta(p^2-m^2),\\
\label{Mperp-1-reduce}
& &p^\nu \nabla_\nu \left[{M}^{(1)\mu}_{\perp} \delta(p^2-m^2) \right]\nonumber\\
&=&-(\frac{p_n}{\bar p^2} \bar p^\mu E^\nu_\perp +  \bar\epsilon^{\mu\nu\alpha}B_\alpha)
{M}^{(1)}_{\perp\nu} \delta(p^2-m^2)\nonumber\\
& &-\frac{ m p_n }{\bar p^2}E^\mu_\perp \sum_{s} s{J}^{(1)}_{sn}\delta(p^2-m^2).
\end{eqnarray}
If we choose the trivial solution
\begin{eqnarray}
\label{Specific-a}
{J}^{(1)}_{sn} = 0, \ \ \
{M}^{(1)\mu}_{\perp}= 0,
\end{eqnarray}
we  obtain the  specific solution for $\mathcal{J}^{(1)}_{sn} $ and ${\mathcal{M}}^{(1)\mu}_{\perp}$
\begin{eqnarray}
\label{Specific-b}
\mathcal{J}^{(1)}_{sn} &=& \frac{s}{2p_n}\tilde\Omega^{\nu\lambda}p_\nu n_\lambda \mathcal{J}^{(0)\prime}_{sn},\nonumber\\
{\mathcal{M}}^{(1)\mu}_{\perp}&=&
\frac{ m n_\lambda }{2p_n\bar p^2}\left(\bar p^2 \tilde\Omega^{\mu\lambda}  - \bar p^\mu \tilde\Omega^{\nu\lambda}\bar p_\nu  \right)
\sum_{s} \mathcal{J}_{sn}^{(0)\prime}.
\end{eqnarray}
It follows that all  other Wigner functions can be calculated directly:
\begin{eqnarray*}
{ \mathscr{J}}_{s}^{(1)\mu}
&=&-\frac{s}{2}  \tilde\Omega^{\mu\nu}p_\nu  \mathcal{J}_{sn}^{(0)\prime} \delta(p^2-m^2)\nonumber\\
& & + s \tilde F^{\mu\nu}p_\nu  \mathcal{J}_{sn}^{(0)} \delta'(p^2-m^2),\\
\label{M-1-EQ-2}
{\mathscr{M}}^{(1)\mu} &=&\frac{m}{2}\tilde\Omega^{\mu\lambda}n_\lambda\sum_s \mathcal{J}^{(0)\prime}_{sn}\nonumber\\
& & -m \tilde{F}^{\mu\lambda}n_\lambda \sum_s \mathcal{J}^{(0)}_{sn}\delta'(p^2-m^2),\\
{\mathscr{S}}^{(1)\mu\nu}
&=&m\Omega^{\mu\nu}\mathcal{J}_{sn}^{(0)\prime}\delta(p^2-m^2) \nonumber\\
& &- 2 m  F^{\mu\nu} \mathcal{J}^{(0)}_{sn}\delta'(p^2-m^2),\\
\tilde{\mathscr{S}}^{(1)\mu\nu}
&=&m\tilde\Omega^{\mu\nu}\mathcal{J}_{sn}^{(0)\prime}\delta(p^2-m^2) \nonumber\\
& &- 2 m \tilde F^{\mu\nu} \mathcal{J}^{(0)}_{sn}\delta'(p^2-m^2),\\
\label{P-1-EQ-2}
\mathscr{P}^{(1)} &=& 0,\\
\label{F-1-EQ-2}
\mathscr{F}^{(1)} &=& 0.
 \end{eqnarray*}
We note that the Wigner functions ${ \mathscr{J}}_{s}^{(1)\mu}$, ${\mathscr{S}}^{(1)\mu\nu}$ and $\tilde{\mathscr{S}}^{(1)\mu\nu}$
are all independent of the auxiliary vector $n^\mu$,  as expected.
\section{General solution}
\label{sec:General}

In the previous section, we chose specific solution (\ref{Specific-a}) or (\ref{Specific-b}).
It is important to investigate whether there
 are any other possible solutions, or what the general solution is. To do this, we must express the functions ${J}^{(1)}_{sn}$ and ${M}^{(1)\mu}_{\perp}$ in terms of all kinds of possible vectors or tensors involved in the problem, such as $n^\mu$, $u^\mu$, $F^{\mu\nu}$, $\Omega^{\mu\nu}$, and their derivatives. Hence,  it would be easier to identify $n^\mu$ as $u^\mu$. However, we cannot naively replace $n^\mu$ with $u^\mu$ because $n^\mu$ has been assumed to be a constant vector from beginning, while the fluid velocity
 $u^\mu$ varies with space and time.  We can achieve this by using  the frame dependence of distribution functions.
Setting $\eta^\mu =u^\mu, n^{\prime\mu}=u^\mu$ in the transformation rules (\ref{Jsn-Frame-1}) and (\ref{Mmu-Frame-1}), we obtain
\begin{eqnarray}
& &{J}_{su}^{(1)}\delta(p^2-m^2)\nonumber\\
&=& {J}_{sn}^{(1)}\delta(p^2-m^2)   - \frac{s  m (u\cdot {M}_n^{(1)})}{2p_u}\delta(p^2-m^2),\\
& &{M}_{u}^{(1)\mu}\delta(p^2-m^2)\nonumber\\
&=&{M}_n^{(1)\mu}\delta(p^2-m^2) -  \frac{p^\mu (u\cdot  {M}_n^{(1)})}{p_u}\delta(p^2-m^2).\hspace{1cm}
\end{eqnarray}
Since the solutions (\ref{Specific-b}) automatically satisfy   the constraint (\ref{M-p-1}),  the distribution functions
${J}_{sn}^{(1)}$ and ${M}_n^{(1)\mu}$ must also satisfy this constraint.
It follows that the  ${M}_{u\parallel}^{(1)\mu}$  is related to ${J}_{sn}^{(1)}$ and ${M}_n^{(1)\mu}$ by
\begin{eqnarray}
{M}_{u\parallel}^{(1)\mu} &=& \frac{m \bar p^\mu_u }{\bar p^2_u}  \sum_{s} s \mathcal{J}^{(1)}_{su}\nonumber\\
&=& \frac{m \bar p^\mu_u }{\bar p^2_u}  \sum_{s} s \mathcal{J}^{(1)}_{sn}
- \frac{m^2 \bar p^\mu_u }{ \bar p^2_u p_u} u\cdot {M}_n^{(1)},
\end{eqnarray}
which lead to the relations
\begin{eqnarray*}
{J}_{su}^{(1)} &=& {J}_{sn}^{(1)}
-\frac{s  m }{2p_u}{u\cdot {M}_{n\perp}^{(1)}}\nonumber\\
& & -\frac{s  m^2 (p_u - p_n n\cdot u) }{2p_u\bar p^2 } \sum_{s'}s' {J}_{s'n}^{(1)},\hspace{1.5cm}\\
{M}_{u\perp}^{(1)\mu}&=&{M}_{n\perp}^{(1)\mu}
+ \frac{(p_u p^\mu -m^2  u^\mu) }{p^2-p_u^2}u\cdot {M}_{n\perp}^{(1)}\nonumber\\
& &+\frac{mp_n}{\bar p^2}\left[\frac{(p_n - p_u n\cdot u)\bar p^\mu_u }{p^2-p_u^2} \right.\nonumber\\
& &\left.\hspace{1cm} +  (u^\mu n\cdot u  -n^\mu)\frac{}{} \right] \sum_s s \mathcal{J}^{(1)}_{sn}.
\end{eqnarray*}
For simplicity,  on-shell delta function $\delta(p^2-m^2)$ has been omitted from  both sides of the  above three equations.
Then, from these relations and kinetic equations (\ref{Jsn-1-reduce}) and (\ref{Mperp-1-reduce}), it is straightforward to derive the
following equations for ${J}^{(1)}_{su}$ and ${M}_{u\perp}^{(1)\mu}$:
\begin{eqnarray}
\label{Jsn-eq-1}
& &p^\mu \nabla_\mu\left[{J}^{(1)}_{su}\delta(p^2-m^2)\right]\nonumber\\
&=&\frac{s  m }{2p_u} \left( E^{\mu}- p^\nu \nabla_\nu u_\mu \right) \nonumber\\
& &\times \left({M}^{(1)}_{\perp\mu} + \frac{m\bar p_\mu}{\bar p^2}\sum_{s'} s'{J}^{(1)}_{s'u}\right)\delta(p^2-m^2),\\
\label{Mperp-eq-1}
& &p^\nu \nabla_\nu \left[{M}^{(1)\mu}_{u\perp} \delta(p^2-m^2) \right]\nonumber\\
&=&-\frac{ m p_u }{\bar p^2}\left[E^\mu_\perp -\Delta_\perp^{\mu\nu} p^\lambda (\nabla_\lambda u_\nu) \right]
\sum_{s} s{J}^{(0)}_{su}\delta(p^2-m^2)\nonumber\\
& &-\left(\frac{p_u }{\bar p^2}\bar p^\mu E^\nu + \bar\epsilon^{\mu\nu\alpha}B_\alpha
+\frac{\bar p^2 u^\mu -p_u \bar p^\mu}{\bar p^2}p^\lambda \nabla_\lambda u^\nu\right)\nonumber\\
& &\times {M}^{(1)}_{u\perp\nu}\delta(p^2-m^2),
\end{eqnarray}
where we have defined the transverse projector
\begin{eqnarray*}
\Delta^{\mu\nu}_\perp= \Delta^{\mu\nu} - \frac{1}{\bar p^2}\bar p^\mu \bar p^\nu.
\end{eqnarray*}
In general, we can express ${J}_{su}^{(1)}$ and ${M}_{u\perp}^{(1)\mu}$ as
\begin{eqnarray}
\label{J-1-Eq}
{J}_{su}^{(1)}    &=& \frac{s}{p_u}\tilde{\Omega}^{\mu\nu}p_\mu u_\nu \mathcal{X}_\Omega^J + \frac{s}{p_u^3}\tilde{F}^{\mu\nu}p_\mu u_\nu \mathcal{X}_F^J ,\\
\label{M-1-Eq}
{M}_{u\perp}^{(1)\mu}
&=&\frac{m}{p_u^2}\Delta^\mu_\lambda \tilde{\Omega}^{\lambda\nu} \bar p_\nu \tilde{\mathcal{X}}_\Omega^M
+\frac{m}{p_u^4}\Delta^\mu_\lambda  \tilde{F}^{\lambda\nu} \bar p_\nu \tilde{\mathcal{X}}_F^M\nonumber\\
& &+ \frac{m}{p_u \bar p^2}\left(\bar p^2\tilde{\Omega}^{\mu\nu}  -\bar p^\mu_u \tilde{\Omega}^{\lambda\nu}p_\lambda  \right)u_\nu \mathcal{X}_\Omega^M\nonumber\\
& &+\frac{m}{p_u^3 \bar p^2}\left(\bar p^2\tilde{F}^{\mu\nu}  -\bar p^\mu_u \tilde{F}^{\lambda\nu}p_\lambda  \right)u_\nu \mathcal{X}_F^M.
\end{eqnarray}
Without loss of generality, we can assume all $\mathcal{X}$ are functions of
\begin{eqnarray*}
z & = & \beta\cdot p-\bar{\mu},\ \tilde{z}=\beta\cdot p+\bar{\mu},\ \ \ \bar m= m/T.
\end{eqnarray*}
By substituting  expressions (\ref{J-1-Eq}) and (\ref{M-1-Eq}) into the kinetic equations (\ref{Jsn-eq-1}) and (\ref{Mperp-eq-1})
and requiring the equations to hold, we obtain
\begin{eqnarray*}
\label{Result-1}
\mathcal{X}_F^J   = 0,\ \ \  \mathcal{X}_F^M = 0,\ \ \ {\tilde{\mathcal{X}}_F^M = 0}.
\end{eqnarray*}
The necessity of these conditions arises from an imbalance: the field-derivative terms on the left-hand side have no counterparts on the right-hand side to cancel them.  Note that the assumption of a varying electromagnetic field is  essential for this simple  conclusion, as the derivative terms  automatically vanish for a constant electromagnetic field.
Now let us determine the remaining terms associated with vorticity tensor from the kinetic equations in global equilibrium
\begin{eqnarray*}
& &p^\lambda \nabla_\lambda\left[\frac{s}{p_u}\tilde{\Omega}^{\mu\nu}p_\mu u_\nu \mathcal{X}_\Omega^J\delta(p^2-m^2)\right]\nonumber\\
&=&\frac{s  m^2}{2p_u^2\bar p^2} \left[ E_{\mu}- p^\lambda (\nabla_\lambda u_\mu)\right]
\left[\left(\bar p^2\tilde{\Omega}^{\mu\nu}  -\bar p^\mu \tilde{\Omega}^{\kappa\nu}p_\kappa  \right)u_\nu \mathcal{X}_\Omega^M
\right.\nonumber\\
& &\left. +2\bar p^\mu\tilde{\Omega}^{\lambda\nu}p_\lambda u_\nu \mathcal{X}_\Omega^J\right]\delta(p^2-m^2)\nonumber\\
& &{+ \frac{s  m^2}{2p_u^3} \left[ E_{\mu}- p^\lambda (\nabla_\lambda u_\mu)\right]\tilde{\Omega}^{\mu\nu}\bar p_\nu
\tilde{\mathcal{X}}_\Omega^M\delta(p^2-m^2)}.
\end{eqnarray*}
After acting the operator $p^\lambda \nabla_\lambda$ on the following terms together with the  decomposition
\begin{eqnarray*}
T\tilde \Omega_{\mu\nu}&=& \omega_\mu u_\nu -\omega_\nu u_\mu +\epsilon_{\mu\nu\rho\sigma}\varepsilon^\rho u^\sigma,
\end{eqnarray*}
and combining like terms, we obtain
\begin{eqnarray}
\label{Equation}
0&=&C_1(E\cdot p)(\omega\cdot p)\delta(p^2-m^2)\nonumber\\
& &+C_2 (\omega\cdot E) \delta(p^2-m^2)\nonumber\\
& &+C_3 (\varepsilon\cdot p)(\omega \cdot p)\delta(p^2-m^2)\nonumber\\
& &+C_4 (\varepsilon\cdot \omega)\delta(p^2-m^2)\nonumber\\
& &+\frac{1}{p_u T}\bar\epsilon_{\lambda\mu \nu} p^\lambda\omega^\mu B^\nu    \mathcal{X}_\Omega^J\delta(p^2-m^2)\nonumber\\
& & -\frac{ m^2}{2p_u^3T} \bar\epsilon_{\lambda\mu\nu}\bar p^\lambda \varepsilon^\mu  E^{\nu}
 \tilde{\mathcal{X}}_\Omega^M \delta(p^2-m^2),
\end{eqnarray}
where the coefficients $C_1$, $C_2$, $C_3$, $C_4$ read
\begin{eqnarray*}
C_1 &=&\frac{1}{p_u^2 T}\left[ \mathcal{X}_\Omega^J - \frac{2 p_u}{T}  \frac{\partial\mathcal{X}_\Omega^J}{\partial {\tilde z}}
-\frac{m^2}{2\bar p^2}(2\mathcal{X}_\Omega^J - \mathcal{X}_\Omega^M)\right]\\
C_2 &=&\frac{1}{T}\left( \mathcal{X}_\Omega^J -\frac{ m^2}{2p_u^2} \mathcal{X}_\Omega^M\right)  \\
C_3 &=&\frac{m}{p_u T}\left[ \frac{1}{T} \frac{\partial\mathcal{X}_\Omega^J}{\partial {\bar m}}
-\frac{ m}{2\bar p^2}  (2\mathcal{X}_\Omega^J - \mathcal{X}_\Omega^M)
+ \frac{ m}{2p_u^2}\tilde{\mathcal{X}}_\Omega^M \right] \\
C_4 &=&\frac{m^2T}{2p_u} \left( 2\mathcal{X}_\Omega^J -\mathcal{X}_\Omega^M
-\frac{\bar p^2}{p_u^2}\tilde{\mathcal{X}}_\Omega^M\right).
\end{eqnarray*}
Given that the first four terms are linearly independent, we obtain the constraints below:
\begin{eqnarray*}
C_1 = 0,\ \ \ C_2 = 0,\ \ \
C_3 = 0,\ \ \ C_4 = 0,
\end{eqnarray*}
which lead to
\begin{eqnarray}
\label{require}
\mathcal{X}_\Omega^J = \frac{ m^2}{2p_u^2} \mathcal{X}_\Omega^M,\
\tilde{\mathcal{X}}_\Omega^M = \mathcal{X}_\Omega^M ,\
\frac{\partial\mathcal{X}_\Omega^J}{\partial {\tilde z}} = 0,\
\frac{\partial\mathcal{X}_\Omega^J}{\partial {\bar m}} = 0.
\end{eqnarray}
Once these relations are satisfied,  Eq.(\ref{Equation}) becomes
\begin{eqnarray}
\label{Equation-1}
0&=& \frac{1}{p_u T}\bar\epsilon_{\lambda\mu \nu} p^\lambda\omega^\mu B^\nu    \mathcal{X}_\Omega^J\delta(p^2-m^2)\nonumber\\
& &- \frac{ m^2}{2p_u^3T} \bar\epsilon_{\lambda\mu\nu} p^\lambda \varepsilon^\mu  E^{\nu} \tilde{\mathcal{X}}_\Omega^M \delta(p^2-m^2)
\end{eqnarray}
Using the constraint condition (\ref{con-tF-a})
\begin{eqnarray*}
\epsilon_{\mu\nu\rho\sigma}u^{\nu}\left(\varepsilon^{\rho}E^{\sigma}-\omega^{\rho}B^{\sigma}\right) & = & -u^{\nu}u^{\lambda}\partial_{\lambda}\tilde{F}_{\mu\nu}\label{dtF-B-omega}
\end{eqnarray*}
and substituting it into Eq.(\ref{Equation-1}), we obtain
\begin{eqnarray}
\label{require-f}
\frac{1}{p_u T}u^{\nu}  p^\mu u^{\lambda}(\partial_{\lambda}\tilde{F}_{\mu\nu}) \mathcal{X}_\Omega^J\delta(p^2-m^2)
&=& 0
\end{eqnarray}
Evidently, the only solution satisfying Eqs. (\ref{require}) and (\ref{require-f}) for an arbitrarily varying electromagnetic field is
\begin{eqnarray*}
\label{Result-2}
\mathcal{X}_\Omega^J =0 ,\ \ \ \ \mathcal{X}_\Omega^M=0,\ \ \ \ \tilde{\mathcal{X}}_\Omega^M=0
\end{eqnarray*}
The results (\ref{Result-1}) and (\ref{Result-2}) demonstrate that the particular  solution (\ref{Specific-b}) is unique.
We have therefore uniquely determined the GCKT under  global equilibrium  conditions. The essential point is the  introduction of a varying electromagnetic field. Without it,  Eq.(\ref{require-f})
would be automatically satisfied for a constant  field,  thereby allowing all terms compatible with  the relations (\ref{require}).  This explains the discrepancy between our previous results
in \cite{Ma:2022ins} and \cite{Gao:2019znl}
where a constant electromagnetic field was assumed. It is thus  remarkable that  the quantum kinetic equations determine the Wigner functions at global equilibrium, it is the presence of a varying electromagnetic field that uniquely fixes their form up to first order of $\hbar$.
\section{Summary}
\label{sec:summary}
In this work, we have systematically investigated the frame dependence of distribution functions within the GCKT. Motivated by the ambiguity in transforming distribution functions across different frames and the non-uniqueness of global equilibrium solutions in prior studies, we derived explicit transformation rules governing this frame dependence. By applying these rules and imposing  a varying electromagnetic field, we were able to uniquely determine the Wigner functions describing fermions of arbitrary mass in global equilibrium under the influence of vorticity and electromagnetic fields. By incorporating the previously neglected frame dependence, our results resolve the previously encountered ambiguities and establish a self-consistent theoretical framework for describing quantum transport phenomena. It thus provides a novel method to determine transport properties in global equilibrium directly from quantum kinetic theory; second and establishes a self-consistent theoretical framework for future applications in relativistic heavy-ion collisions and other related fields.

{\bf{Acknowledgments}}

This work was supported in part by  the National Natural
Science Foundation of China  under Grant
Nos. 12575147, 12321005, and 12175123.

\end{document}